\documentclass[pra,aps,showpacs,twocolumn,superscriptaddress, floatfix]{revtex4-1}
\usepackage{graphicx}
\usepackage[ansinew]{inputenc}
\usepackage{array}
\usepackage{color}
\usepackage{amsmath}
\usepackage{amsxtra}
\usepackage{amstext}
\usepackage{amssymb}
\usepackage{latexsym}

\newcommand{\vv}[1]{\mathbf{#1}}
\newcommand{\vvb}[1]{\boldsymbol{#1}}
\newcommand{\bra}[1]{\left<#1\right|}
\newcommand{\ket}[1]{\left|#1\right>}

\usepackage{cancel,ifthen}
\usepackage[normalem]{ulem}

\newcommand{\comment}[2][NoInPuT]{\ifthenelse{\equal{#1}{NoInPuT}}{}{{\color{blue}\sout{#1}}}{\color{red} #2}}

\begin{document}
\title{Single-atom quantum control of macroscopic mechanical oscillators}
\date{\today}
\author{F. Bariani}
\affiliation{Department of Physics, College of Optical Sciences and B2 Institute, University of Arizona, Tucson, Arizona 85721, USA}
\author{J. Otterbach}
\affiliation{Physics Department, Harvard University, Cambridge, Massachusetts 02138, USA}
\author{Huatang Tan}
\affiliation{Department of Physics, College of Optical Sciences and B2 Institute, University of Arizona, Tucson, Arizona 85721, USA}
\affiliation{Department of Physics, Huazhong Normal University, Wuhan 430079, China}
\author{P. Meystre}
\affiliation{Department of Physics, College of Optical Sciences and B2 Institute, University of Arizona, Tucson, Arizona 85721, USA}
\begin{abstract}
We investigate a hybrid electro-mechanical system consisting of a pair of charged macroscopic mechanical oscillators coupled to a small ensemble of Rydberg atoms. The resonant dipole-dipole coupling between an internal atomic Rydberg transition and the mechanics allows cooling to its motional ground state with a single atom despite the considerable mass imbalance between the two subsystems. We show that the rich electronic spectrum of Rydberg atoms, combined with their high degree of optical control, paves the way towards implementing various quantum-control protocol for the mechanical oscillators.  
\end{abstract}
\maketitle

Hybrid quantum systems are attracting increasing attention as fundamental building blocks for applications requiring the manipulation of quantum states, such as e.g. quantum information science and quantum metrology \cite{wallquist_hybrid,qinterface,hybrid_sc}. These systems come in many variations, including, e.g., ultracold atoms coupled to photons in high-$Q$ optical cavities or in optical lattices, superconducting qubits coupled to microwave fields,  NV centers coupled to photonic-crystal cavities, single atoms, artificial atoms, or photons coupled to mechanical oscillators, and many more. In this context, Rydberg atoms have been proposed to realize quantum interfaces with superconducting qubits or as a direct means to manipulate the spatio-temporal properties of photons\cite{rydberg_om_sc,hafezi_rydberg,petrosyan_rydberg,petrosyan_rydberg_2, Gorshkov2011, Otterbach2013}. However, the realization of strong and scalable interactions in free space presents a formidable task. A promising technology to overcome this obstacle is offered by quantum interfaces based on electro-mechanical and magneto-mechanical forces that can result in strong coupling in free space ,without the need for cavity mediated enhancement of the interaction \cite{rabl09,bennet_em,rabl10,jun_om,swati_om}.

The rapid developments in quantum optomechanics provides attractive possibilities to tackle this challenge, by allowing to couple microscopic to macroscopic systems operating deep in the quantum regime \cite{kip_vah_om,aspel_om,mar_gir_om,aspel_mey_schw_om,meystre_om,stamper_om,review_om}. Here we propose a scheme that exploits the remarkable properties of Rydberg atoms to dipole-couple one or more macroscopic mechanical oscillators to a small atomic ensemble in free space.

The strong electric dipole-dipole coupling between the Rydberg atom and the mechanics offers a number of distinct advantages \cite{gallagher_book}. First, it allows to implement a recycling scheme to cool the mechanics with a single atom, despite the large mass imbalance between the two subsystems. Second, it is possible to select Rydberg transitions such that the frequencies of the atomic transition and the mechanics are either perfectly matched, or adjusted precisely to enhance or inhibit specific aspects of the electro-mechanical coupling. Third, Rydberg level lifetimes are very long, allowing to operate in a regime where dissipation is negligible, thus enabling the coherent manipulation of the quantum state of the mechanics. Finally, using Rydberg-blockade interactions allows one to enhance the atom-photon cross-section. This makes it feasible to extend these properties to the single photon regime, realizing and ideal interface between single photons and single phonons. To illustrate these features we show how to generate large Fock states and number superposition states of a cantilever, as well as entangled states of a pair of oscillators.  

{\em The system --} To set the stage we consider a hybrid system composed of two electrically charged tuning fork-shaped cantilevers coupled to a single Rydberg atom, see Fig.~1. Electric charges $\pm Q$ separated by an oscillating distance $d(t)$ are located at the opposite tips of the cantilevers, and the atom is trapped in its motional ground state. The trap, with characteristic size $a_{\rm tr}$ located at a distance $R$ half-way between the cantilevers, with $R \gg (a_{\rm tr}, d)$ and we assume that it operates in the Lamb-Dicke regime. In this limit its position is fixed on the time scales considered in this work and recoil effects are negligible. Furthermore we assume that it can also trap the atom when excited in the Rydberg states of interest \footnote{See details in the Supplemental Material which includes Refs. \cite{saffmantrap,kuzmichtrap,saffmantrap2, honer_deph, Kubler2013, bariani_retrieval, myro_molmer, petrosyan_2013}}.   
\begin{figure}[]
\begin{center}
\includegraphics[width = \columnwidth]{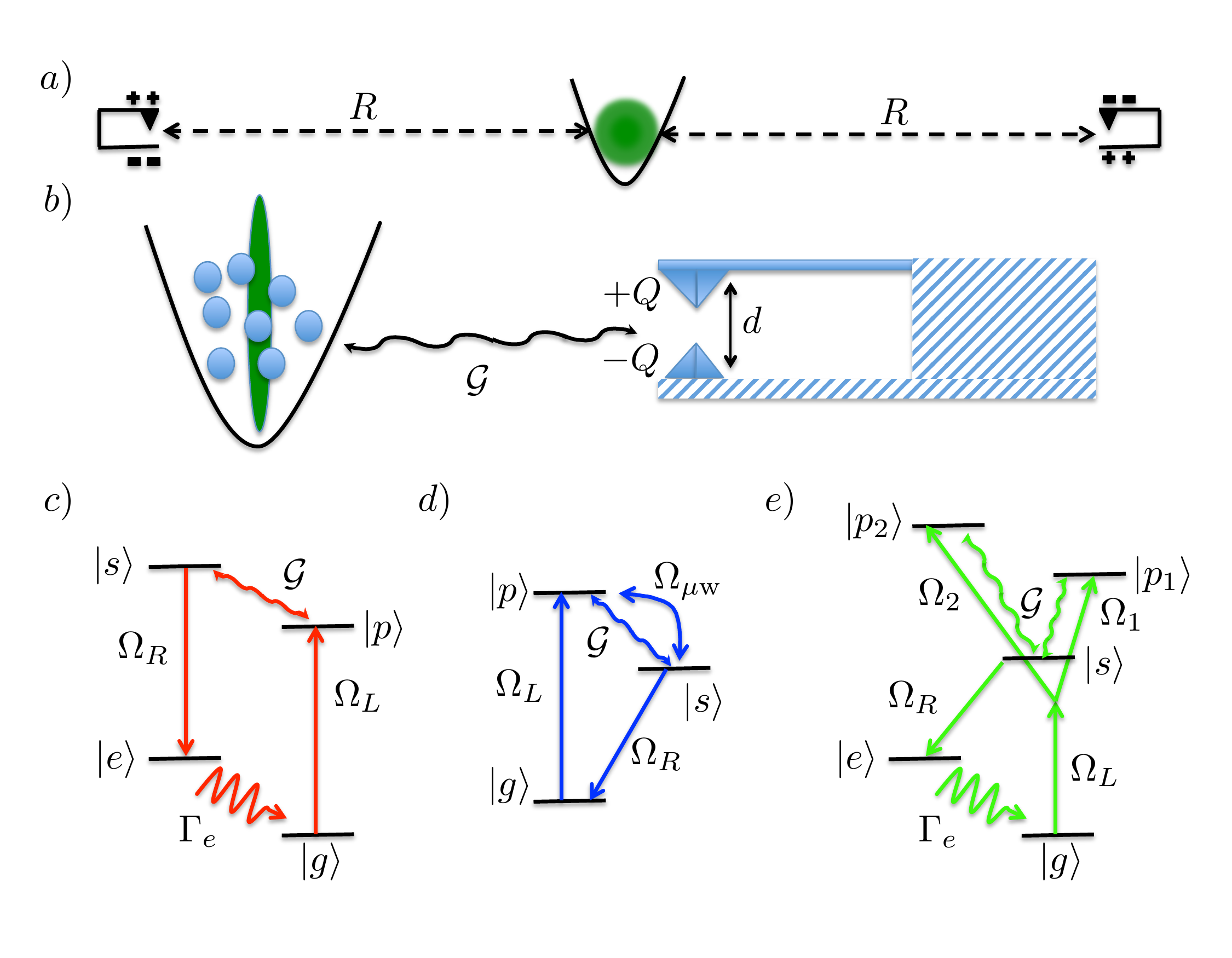}
\caption{(Colors online). a) Geometry of the atom-double cantilever system. b) Structure of the charged cantilever coupled to the blockaded ensemble containing a Rydberg elliptical excitation. c) Atomic energy levels and external fields for ground state cooling of mechanical resonator.  d) Atomic energy levels and external fields for engineering of mechanical quantum state. e) Atomic structure to generate highly entangled cantilever state.}
\end{center}
\end{figure}
This system is described by the Hamiltonian
\begin{equation}
H = H_c + H_{\rm at} + V + H_{\rm loss}.
\label{eq:totalH}
\end{equation}
Here 
\begin{equation}
H_c = \sum_{\ell = 1,2} \hbar \omega_\ell \hat b_\ell^{\dagger} \hat b_\ell
\end{equation}
accounts for the free evolution of the quantized mode of motion of the mechanical oscillator $\ell$ of frequency $\omega_\ell$, with bosonic phonon annihilation and creation operators $\hat b_\ell$ and $\hat b_\ell^\dagger$,
\begin{equation}
H_{\rm at} = H_0 +  H_{\rm tr} + H_{\rm ext} 
\end{equation}
where $H_0$ describes the relevant electronic energy levels of the atom,  $H_{\rm tr}$ its trapping potential, and  $H_{\rm ext}$ its interaction with one or more optical and microwave fields, depending on the specific application being considered. $H_{\rm loss}$ accounts for dissipation and decoherence due to coupling to external reservoirs, and   $V$ describes the interaction between the atom and the mechanics. It can be decomposed into two contributions, $V = \sum_\ell V_{\ell,\rm at} + V_{12}$, $\ell=\{1,2\}$, where the first  term accounts for the electric dipole coupling between the cantilever and the atom
\begin{equation}
V_{\ell,\rm at} = - \sum_{\ell} \vv{E}_\ell(t) \cdot \vvb{\mu}_{\rm at},
\label{eq:Hlat}
\end{equation}
with ${\bf E}_\ell(t)$ the electric field generated by cantilever $\ell$ at the location of the atom and $\vvb{\mu}_{\rm at}$ the dipole moment of the atomic transition under consideration~\footnote{We use bold symbols to indicate vectors, while the italic refers to their moduli}. The third term $V_{12}$, capturing the cantilever-cantilever coupling, is negligible, as briefly discussed later on. 

The dominant contribution to  ${\bf E}_\ell(t)$ is the dipole term, proportional to $\vvb{\mu}_\ell = Q\vv{d}_\ell$. For cantilevers aligned along the $z$-axis and located at $z_\ell \gg d$ from the atomic trap center we have \cite{jackson}
\begin{equation}
\vv{E}_\ell = \frac{1}{4\pi\epsilon_0} \frac{Q}{z_\ell^{3}} \left[\vvb{\mu}_\ell - 3\frac{\vv{z}_\ell (\vv{z}_\ell\cdot \vvb{\mu}_\ell)}{z_\ell^2} \right].
\end{equation}
The static part of the potential (\ref{eq:Hlat}) mixes opposite-parity states while preserving the total angular momentum along the direction connecting the two dipoles. To cancel the static field acting on the atom, we choose the geometry of Fig.~1a, where the two cantilevers have opposite dipoles, thus preserving the atomic non-interacting energy level structure. (We may still allow for a small imbalance in charge between the cantilevers in order to choose a polarization axis for the atomic dipole.) 
 
The time-dependent part of the electric field, resulting from the oscillatory motion of the cantilevers, drives the atomic Rydberg transition. Assuming the displacement to be aligned with the $x$-axis we have $d_\ell(t) = d_\ell + \hat x_\ell(t)$. In terms of phononic annihilation and creation operators $\hat x_\ell = x_{\rm zp,\ell} (\hat b_\ell + \hat b_\ell^{\dagger})$ where $x_{\rm zp,\ell} = \sqrt{\hbar/(2 m_{\rm eff, \ell} \omega_\ell)}$ is the zero point motion of the resonator mode with effective mass $m_{\rm eff,\ell}$ and frequency $\omega_\ell$. 

In the following we assume that the cantilever frequency $\omega_\ell$ is resonant with a single dipole-allowed Rydberg $ \vert s\rangle-\vert p\rangle$ transition, and that all other transitions can be neglected. To first order in $(x_{{\rm zp},\ell}/d)$, the atom-cantilever coupling reduces then to 
\begin{equation}
V_{\ell,\rm at} = \hbar \mathcal{G}_\ell \left(b_\ell + \hat b_\ell^{\dagger} \right) \left(\hat \sigma_{sp} + \hat \sigma_{ps} \right),
\label{eq:CantileverAtomCoupling}
\end{equation}
where $\hbar \mathcal{G}_\ell = Q x_{\rm zp,\ell} \mu_{sp}/(4\pi\epsilon_0 R^{3})$ with $\mu_{sp}$ being the dipole matrix element of the $ \vert s\rangle-\vert p\rangle$ transition, and $\hat \sigma_{ab} = \ket{a}\bra{b}$. Within the rotating-wave approximation (RWA) Eq.~(\ref{eq:CantileverAtomCoupling}) reduces to the Jaynes-Cummings interaction. Note that by tuning the charge on the cantilever, it is feasible to enter the ultra-strong coupling regime $\mathcal{G}_\ell \gg \omega_\ell$ in which case the RWA breaks down and we have to solve the full Rabi model. This regime could be used to study, e.g., spontaneous phonon generation from vacuum in analogy with the dynamical Casimir effect \cite{beyondRWA}.

Returning to the cantilever-cantilever coupling $V_{12}$ we have
\begin{align}
V_{12} =  \frac{Q^2}{4 \pi \epsilon_0} \frac{1}{8 R^3} & \left[d^2 + d x_{\rm zp,1} (\hat b_1 + \hat b_1^{\dagger}) + d x_{{\rm zp},2} (\hat b_2 + \hat b_2^{\dagger}) \right.\nonumber \\
 & \left. + x_{{\rm zp},1} x_{{\rm zp},2} (\hat b_1+ \hat b_1^{\dagger}) (\hat b_2 + \hat b_2^{\dagger}) \right]. 
\end{align}
The term proportional to $d^2$ is a global energy shift, the second and third terms may be absorbed in the definition of the displacements and the phonon exchange between the cantilevers is negligible compared to the atom-cantilever interaction and will be ignored in the following \footnote{See Supplemental Material for a detailed comparison of the two terms}.

We can extend these results to the case of a small ensemble of $N \sim 100$ trapped atoms, in which case the atomic Hamiltonian becomes $H_{\rm at} \rightarrow \sum_j H_{{\rm at},j} + \sum_{j > k} V_{jk}$, where the inter-atomic coupling $V_{jk}$ can account either for non-resonant Van der Waals interactions, $\sim 1/r_{jk}^6$, where $r_{jk}$ is the distance between atoms $j$ and $k$, or F{\"o}rster processes, $\sim 1/r_{jk}^3$ \cite{saffman_review,gallagher_book}. If all atoms are initially in their ground electronic state $\ket{g}$ and we operate in Rydberg-blockade regime, $\hbar \Omega_{L,R} \ll (C_{\alpha}/r^{\alpha}_{jl})$, $\forall \{j,l\}$,  $\alpha = \{3, 6\}$,  \cite{blockade_lukin,blockade_tong} we can describe the atomic ensemble as a ``super-atom'' whose states are collective excitations (spin waves) of the form
\begin{equation}
\ket{\Psi_a} = \frac{1}{\sqrt{N}} \sum_{j=1}^{N} e^{i \vv{k}_{\Psi} \cdot \vv{r}_j} \hat \sigma^{j}_{ag} \ket{G}.
\end{equation}
Here $a$ labels the relevant excited electronic state and the many-body ground state $\ket{G}=\ket{g_1\dots g_N}$ denotes the collective ground state. We note that in this case the optical coupling of$\ket{G}$ to some excited electronic state is enhanced by a factor $\sqrt{N}$, see e.g. \cite{kuzmich_blockade}. In the limit $k_\Psi a_\text{tr}\ll 1$ the super-atom can then be treated as a single atom with effectively larger atom-photon cross-section, enabling a strong interaction with single photons \footnote{See details and references in the Supplemental Material.}.

With recent progress in the nanofabrication of cantilevers from single-crystal diamond, mechanical oscillators of very small size and large stiffness have become available~\cite{diamondcantilever, diamondcantilever2}. For a clamped beam of dimensions $(l,w,t) = (0.5,0.05,0.05)\mu$m,  with a Young's modulus $E = 1000$ GPa and a density $\rho = 3\times10^{-3}$ Kg$\cdot$cm$^{-3}$, we obtain $\omega_\ell/2\pi = 3.516\, (t/l^2) \sqrt{E/(12 \rho)} = 578$ MHz \cite{rmpcantilever},  $m_{\rm eff}  = 1.9\times10^{-18}$Kg, and $x_{\rm zp} = 1.7\times 10^{-13}$ m $= 1.1 \times 10^{-3} a_0$, where $a_0$ is the Bohr radius. For the atomic system, consider a rather generic $\ket{p} = \ket{180,p,3/2} \leftrightarrow\ket{s} = \ket{180,s,1/2}$ Rydberg transition in $^{87}$Rb, with transition frequency $\omega/2\pi = 578$ MHz and dipole moment $\mu_{\rm at} \approx 35250\,e a_0$~\cite{kaulakys}.
For these parameters, a single charge on the cantilevers and $R = 5 \mu$m results in $\mathcal{G_\ell}/2\pi = e\, x_{\rm zp,\ell} \mu_{\rm at}/(4\pi\epsilon_0 R^{3} h) = 314$ Hz, a value increased to about 1 MHz for $Q \sim 3\cdot 10^3 \, e$. 

The dominant decay and decoherence mechanisms are the radiative decay of the atom, with a typical radiative linewidth of the order of $\Gamma_{s,p}/2\pi \sim 10$ KHz for Rydberg atoms with principal quantum number $n \approx 100$, and scaling as $n^{-3}$. For the mechanics, quality factors of order $Q \approx 10^6$ are achievable, corresponding to $\Gamma_m/2\pi = 578$ Hz. The heating rate of the mechanical oscillator is independent of the mechanical frequency, $\Gamma_{m,T} = k_BT/(\hbar Q) = 2$ KHz for $T = 100$ mK.

{\em Applications --} In the remainder of this paper we discuss potential applications of this hybrid system, considering first cantilever cooling, and then several examples of quantum engineering of specific phononic states. 

For cantilever cooling we consider the 4-level subsystem of Fig.~1c, with the atom interacting resonantly with cantilever 1, i.e. $\omega_1 = \omega_{sp} \neq \omega_2$, while the interaction with cantilever 2 is off-resonant and can be neglected. The (super-)atom is driven by two classical laser fields of Rabi frequencies $\Omega_i$, frequencies  $\omega_i$ and wave vectors ${\bf k}_i$, $i=\{ R, L\}$, see Fig.~1c, so that
\begin{equation}
H_{\rm ext} = \frac{1}{2\hbar} \left [ \Omega_L \hat \sigma_{gp} e^{i (\omega_L t - \vv{k}_L \cdot \vv{r})} 
+ \Omega_R \hat\sigma_{es} e^{i (\omega_R t- \vv{k}_R \cdot \vv{r})}  + {\rm h.c.} \right ].
\end{equation}
We concentrate on the realistic situation $\Omega_L,\Omega_R, \Gamma_e \gg \mathcal{G}_1$, and neglect the radiative decay of the Rydberg-transition. This leads to a separation of time scales allowing us to describe the dynamical effects of $H_{\rm ext}$ and $V_{\ell,\rm at}$ separately. 

Intuitively, the cooling process can then be understood as follows: For atoms initially in the ground state $|g\rangle$ the resonant driving field $\Omega_L$ nearly saturates the $\ket{g}-\ket{p}$ transition, resulting in $\left<\hat \sigma_{gg}\right> = \left<\hat \sigma_{pp}\right> \approx 1/2$, while the populations of the other atomic states remains negligible. The weak coupling ${\cal G}_1$ between the cantilever and the excited Rydberg states then induces transitions from level $|p\rangle$ to $|s\rangle$, with absorption of a phonon,  at which point  the strong field $\Omega_R$ induces a rapid transition down to level $|e\rangle$, preventing the transfer of energy back to the cantilever. Spontaneous decay from level $|e\rangle$ to $|g\rangle$ finally resets the system and initiates the next cycle in the cooling process. 

Assuming that the atomic populations remain in quasi steady state, one can estimate the effective damping rate of the cantilever phonon population to be of order $\mathcal{G}_1^2/\Omega_L \gg \gamma_m$. This indicates that the cantilever mode may be cooled to its ground state via interaction with even a single Rydberg atom. This is confirmed by a direct numerical solution of the associated master equation~\cite{carmicheal}, as illustrated in Fig.~2 \footnote{Explicit form of the equation of motion may be found in the Supplemental Material}.

\begin{figure}[]
\begin{center}
\includegraphics[width = \columnwidth]{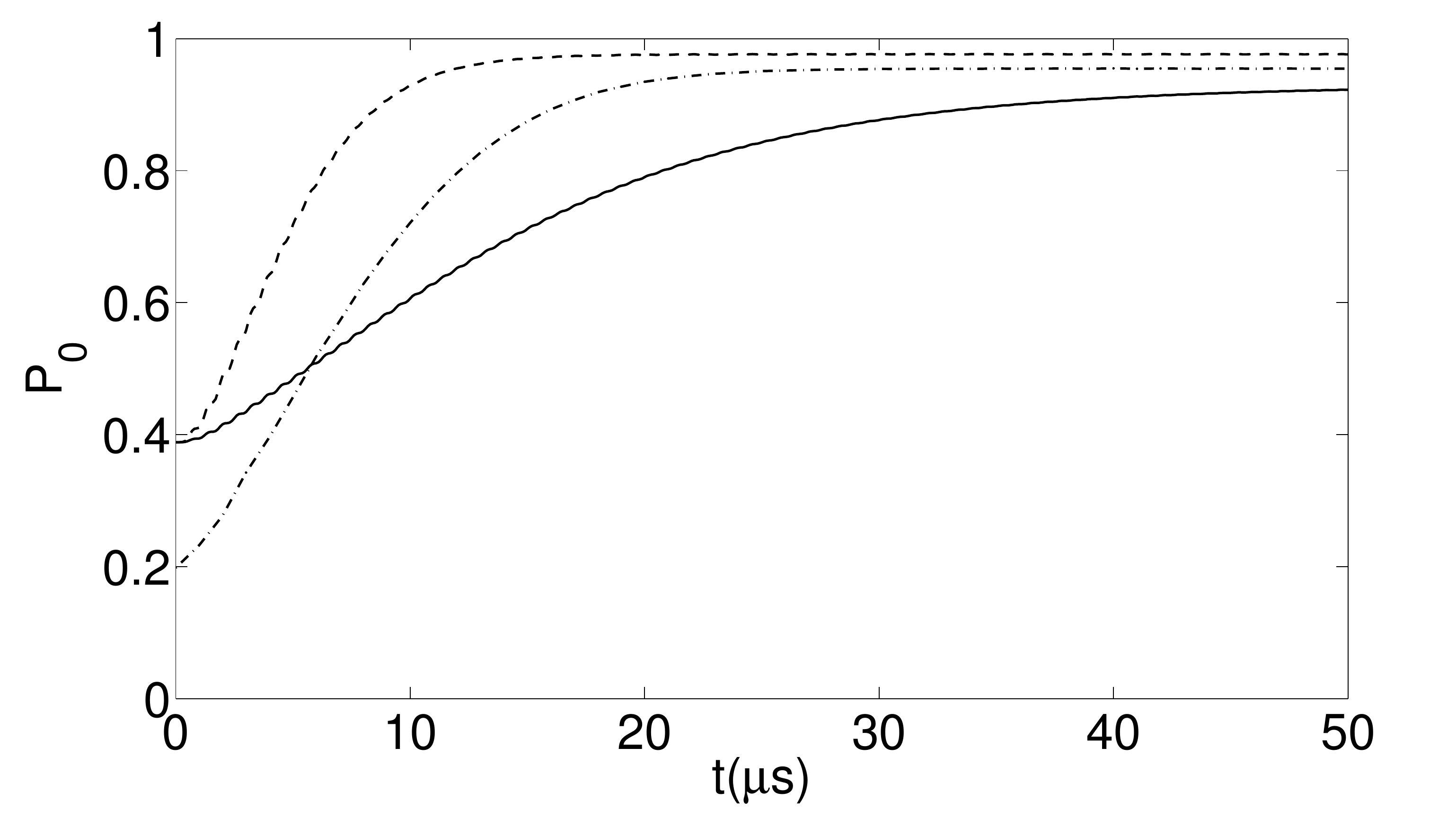}
\caption{Cooling of the mechanical motion via atomic dissipation for $\mathcal{G}/2\pi = 1$MHz (solid line) and $\mathcal{G}/2\pi = 2$MHz (dashed and dot-dashed lines). We plot the population of the mechanical ground state for $\Omega_L = \Omega_R = 2\pi \times 10$ MHz and $\Gamma_e/2\pi = 5$ MHz. The initial temperature is $T = 0.1$ K (solid and dashed lines) and the final effective temperatures are $T_\text{eff} = 0.016$ K (solid) and $T_\text{eff} = 0.011$ K (dashed). For the dot-dashed line: $T = 0.2$ K and $T_\text{eff} = 0.013$ K. The cantilever frequency corresponds to $T_\text{osc}=\hbar\omega_\ell/k_B = 0.028$ K.}
\label{fig:cooling}
\end{center}
\end{figure}

Once the cantilever is in its ground state, coupling to a Rydberg atom can be used to generate arbitrary phonon Fock states. A convenient way to achieve this uses the three-level scheme of Fig.~1d and an interaction of the form
\begin{eqnarray}
H_{\rm ext} &=& \frac{1}{2\hbar} \left[\Omega_L(t)  \hat \sigma_{gp} e^{i (\omega_L t - \vv{k}_L \cdot \vv{r})} + \Omega_R(t)\hat \sigma_{gs} e^{i (\omega_R t- \vv{k}_R \cdot \vv{r})}  \right. \nonumber \\
 &+& \left.  \Omega_{\mu}(t)\hat \sigma_{sp} e^{i (\omega_{\mu} t - \vv{k}_{\mu} \cdot \vv{r})} + {\rm h.c.}\right],
\end{eqnarray}
where the optical fields are now time-dependent pulses, and we have added an additional coupling to a microwave field for future use. 

Starting from the atom in its ground electronic state $|g\rangle$ and the mechanics cooled to $|m=0\rangle$, a possible protocol to prepare an arbitrary phononic Fock state goes as follows: First, excite the atom to the state $\ket{p}$ with a fast $\pi$-pulse $\Omega_L(t)$. The state $\ket{p,m=0}$ is then resonantly coupled to $\ket{s,m=1}$ via the dipole-dipole interaction ${\cal G}_1$, resulting in a perfect exchange of excitation (neglecting dissipation) between the qubit  $\{ |s\rangle , |p\rangle \}$ and the cantilever after an interaction time $\tau_R = \pi/{\cal G}_1$. Finally, the resulting Rydberg state $|p\rangle$ is  coupled back to $|g\rangle$ by a fast $\pi$-pulse $\Omega_R$. By repeating that cycle it is possible to create an arbitrary Fock state $|m\rangle$, provided that its creation time is short compared to the inverse mechanical damping rate. 

\begin{figure}[]
\begin{center}
\includegraphics[width = \columnwidth]{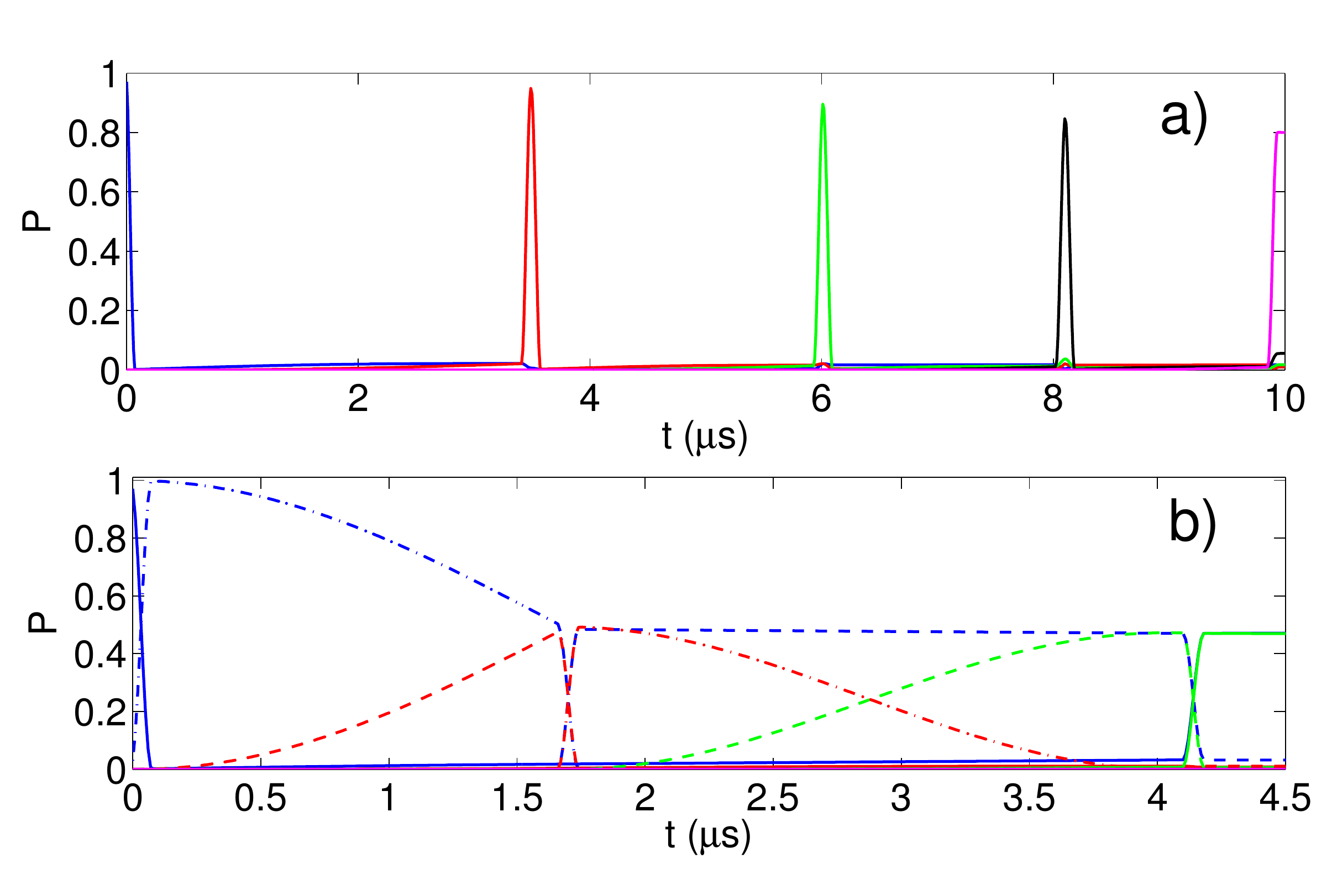}
\caption{(Colors online). a) Generation of a Fock state on the cantilever corresponding to $\ket{\psi_F} = \ket{g,4}$. We show the evolution of the populations of the different phonon number states $\ket{g,m}$ (blue: $m = 0$, red: $m = 1$, green: $m = 2$, black: $m = 3$, magenta: $m = 4$). For the external fields we use square pulses, $\Omega_L/2\pi =  \Omega_R/2\pi = 6$ MHz, while $\mathcal{G}/2\pi = 150$ KHz is constant in time. b) Generation of the mechanical superposition state $\ket{\psi_S} =  (\ket{g,0} - \ket{g,2})/\sqrt{2}$. We use microwave square pulses with $\Omega_{\mu}/2\pi = 6$ MHz. The colors correspond to the different phonon number states as in a), while the line-styles correspond to a specific atomic level: solid = $\ket{g}$, dashed = $\ket{s}$, dot-dashed = $\ket{p}$. The fidelity of the process is $\mathcal{F} =\left| \langle \psi_f|\psi_S \rangle\right|^2 = 0.94 $. Dissipation: atomic states lifetimes are $\Gamma_s/2\pi = \Gamma_p/2\pi = 2$ KHz, and mechanical damping rate is $\Gamma_m/
2\pi = 600$ Hz $(T=0)$.}
\label{fig:merge}
\end{center}
\end{figure}

The coupling between the states $\ket{p,m}$ and $\ket{s,m+1}$ scales as $\sqrt{m+1}$. For $(\Omega_L, \Omega_R) \gg \mathcal{G}_1\sqrt{m+1}$, satisfied even for large $m$ using the above parameters, the total time required for the creation of an $m$-phonon Fock state is therefore approximately $\tau_m = \sum_{j = 1}^{m} (\tau_R/\sqrt{j})$.

Introducing a microwave field of Rabi frequency $\Omega_\mu(t)$ in addition to the pair of optical fields permits to prepare more complex superpositions of phononic Fock states, following the protocol originally proposed by Law and Eberly \cite{law_eberly}. In one such example, the initial excitation of the atom to the $\ket{p}$ state via a $\pi$-pulse $\Omega_L(t)$ is followed by a partial transfer of  excitation from the atom to the cantilever in a time $\pi/(2 \mathcal{G}_1)$, resulting in the superposition [$i(\ket{p,0} + i \ket{s,1})/\sqrt{2} $]. A subsequent microwave $\pi$-pulse exchanges the atomic population in the Rydberg states to [$(-\ket{s,0} - i \ket{p,1})/\sqrt{2} $]. Within the RWA the state $\ket{s,0}$ is not coupled to the cantilever and we observe an excitation transfer from the electronic state $|p\rangle$ to the cantilever in time $\pi/(\sqrt{2} \mathcal{G}_1)$. The resulting state is $(-1/\sqrt{2}) (\ket{s,0} - \ket{s,2})$. Finally the atom is brought back to its ground state via a $\pi$-pulse $\Omega_R(t)$. Note that for the protocol to work, it is important to have  $\Omega_{\mu } \gg \mathcal{G}_1$. This is necessary in order to avoid having to turn off the resonant dipole-dipole coupling while performing the microwave transition.

Figure 3 shows the result of a full numerical simulation of these 2 protocols, including mechanical damping and spontaneous emission. At low phonon occupation dissipation is mainly due to the atomic decay, with losses of order $\Gamma_{s,p} \pi/(\sqrt{m} \mathcal{G}_1)$. As the number of phonon increases, though, mechanical dissipation starts to dominate, as clearly apparent in subsequent cycle of the Fock state generation, see Fig.~3a. For the parameters of the simulations both target states are generated in a few microseconds.

So far, the second cantilever was merely used to balance out the dc component of the electric dipole-dipole interaction. However, it can be used to generate entangled cantilever states \cite{joshi10}. Typically, though, the preparation of more complex quantum states requires increasingly complicated  multilevel atomic schemes. We discuss one example, using the five-level atomic scheme of Fig.~1e, which permits the generation of a NOON state with $N = 2$. Here $\omega_{p_1s} = \omega_1$ and $\omega_{p_2s} = \omega_2$. In this case the couplings from $\ket{g}$ to $\ket{p_1}$ and $\ket{p_2}$ are provided by two-photon processes that combine $\Omega_L$ and respectively $\Omega_1$ and $\Omega_2$. In the first step of the protocol we assume $\Omega_1 = \Omega_2$ and  generate the state $(
\ket{0,p_1,0} + \ket{0,p_2,0})/\sqrt{2}$. This state is then transferred to the state [$(\ket{1,s,0} + \ket{0,s,1})/\sqrt{2}$] via the atom-cantilever interaction. Subsequent dissipation brings the atoms back to the ground state resulting in the state $(\ket{0,g,1} + \ket{1,g,0})/\sqrt{2}$. In the second step, the phase of one of the coupling lasers is inverted, giving, e.g. $\Omega_1 = -\Omega_2$. This generates the state $(\ket{1,p_1,0} - \ket{1,p_2,0} + \ket{0,p_1,1} - \ket{0,p_2,1})/2$, which is coupled via $V_{1,\rm at} + V_{2,\rm at}$ to the state $(\ket{2,s,0} - \ket{0,s,2})/\sqrt{2}$. After transferring the atom back to the ground state we obtain our target state.

In conclusion, we have  analyzed a hybrid system that couples a microscopic system consisting of one Rydberg atom to a pair of charged macroscopic cantilevers. We have derived the resonant dipole-dipole coupling between these systems, and found that it can be quite significant, in the MHz range. The experimental requirements to realize this system are within, or close to, presently available technology. We have  proposed a scheme to achieve ground state cooling of the cantilevers, and discussed examples of quantum state engineering protocols that exploit the combined advantages of this coupling and the excellent coherent control over atomic systems. The spurious effects of dissipation and decoherence are kept manageable due to the long lifetime of Rydberg states and the availability of mechanical oscillators with high $Q$ factor. In the future, we plan to investigate possible schemes to realize quantum state tomography and to exploit the blockade effect to propose a high efficiency single-photon to single-phonon transducer. 

We thank S.K. Steinke, H. Seo, L.F. Buchmann and E.M. Wright for useful discussions. F.B. and P.M. acknowledge funding from the DARPA ORCHID and QuASAR programs through grants from AFOSR and ARO, The US Army Research Office, and NSF.  J.O. acknowledges support from the Harvard Quantum Optics Center.

 \bibliography{references}

\begin{thebibliography}{48}%
\makeatletter
\providecommand \@ifxundefined [1]{%
 \@ifx{#1\undefined}
}%
\providecommand \@ifnum [1]{%
 \ifnum #1\expandafter \@firstoftwo
 \else \expandafter \@secondoftwo
 \fi
}%
\providecommand \@ifx [1]{%
 \ifx #1\expandafter \@firstoftwo
 \else \expandafter \@secondoftwo
 \fi
}%
\providecommand \natexlab [1]{#1}%
\providecommand \enquote  [1]{``#1''}%
\providecommand \bibnamefont  [1]{#1}%
\providecommand \bibfnamefont [1]{#1}%
\providecommand \citenamefont [1]{#1}%
\providecommand \href@noop [0]{\@secondoftwo}%
\providecommand \href [0]{\begingroup \@sanitize@url \@href}%
\providecommand \@href[1]{\@@startlink{#1}\@@href}%
\providecommand \@@href[1]{\endgroup#1\@@endlink}%
\providecommand \@sanitize@url [0]{\catcode `\\12\catcode `\$12\catcode
  `\&12\catcode `\#12\catcode `\^12\catcode `\_12\catcode `\%12\relax}%
\providecommand \@@startlink[1]{}%
\providecommand \@@endlink[0]{}%
\providecommand \url  [0]{\begingroup\@sanitize@url \@url }%
\providecommand \@url [1]{\endgroup\@href {#1}{\urlprefix }}%
\providecommand \urlprefix  [0]{URL }%
\providecommand \Eprint [0]{\href }%
\providecommand \doibase [0]{http://dx.doi.org/}%
\providecommand \selectlanguage [0]{\@gobble}%
\providecommand \bibinfo  [0]{\@secondoftwo}%
\providecommand \bibfield  [0]{\@secondoftwo}%
\providecommand \translation [1]{[#1]}%
\providecommand \BibitemOpen [0]{}%
\providecommand \bibitemStop [0]{}%
\providecommand \bibitemNoStop [0]{.\EOS\space}%
\providecommand \EOS [0]{\spacefactor3000\relax}%
\providecommand \BibitemShut  [1]{\csname bibitem#1\endcsname}%
\let\auto@bib@innerbib\@empty
\bibitem [{\citenamefont {Wallquist}\ \emph {et~al.}(2009)\citenamefont
  {Wallquist}, \citenamefont {Hammerer}, \citenamefont {Rabl}, \citenamefont
  {Lukin},\ and\ \citenamefont {Zoller}}]{wallquist_hybrid}%
  \BibitemOpen
  \bibfield  {author} {\bibinfo {author} {\bibfnamefont {M.}~\bibnamefont
  {Wallquist}}, \bibinfo {author} {\bibfnamefont {K.}~\bibnamefont {Hammerer}},
  \bibinfo {author} {\bibfnamefont {P.}~\bibnamefont {Rabl}}, \bibinfo {author}
  {\bibfnamefont {M.}~\bibnamefont {Lukin}}, \ and\ \bibinfo {author}
  {\bibfnamefont {P.}~\bibnamefont {Zoller}},\ }\href
  {http://stacks.iop.org/1402-4896/2009/i=T137/a=014001} {\bibfield  {journal}
  {\bibinfo  {journal} {Physica Scripta}\ }\textbf {\bibinfo {volume} {2009}},\
  \bibinfo {pages} {014001} (\bibinfo {year} {2009})}\BibitemShut {NoStop}%
\bibitem [{\citenamefont {Hammerer}\ \emph {et~al.}(2010)\citenamefont
  {Hammerer}, \citenamefont {S\o{}rensen},\ and\ \citenamefont
  {Polzik}}]{qinterface}%
  \BibitemOpen
  \bibfield  {author} {\bibinfo {author} {\bibfnamefont {K.}~\bibnamefont
  {Hammerer}}, \bibinfo {author} {\bibfnamefont {A.~S.}\ \bibnamefont
  {S\o{}rensen}}, \ and\ \bibinfo {author} {\bibfnamefont {E.~S.}\ \bibnamefont
  {Polzik}},\ }\href {\doibase 10.1103/RevModPhys.82.1041} {\bibfield
  {journal} {\bibinfo  {journal} {Rev. Mod. Phys.}\ }\textbf {\bibinfo {volume}
  {82}},\ \bibinfo {pages} {1041} (\bibinfo {year} {2010})}\BibitemShut
  {NoStop}%
\bibitem [{\citenamefont {Xiang}\ \emph {et~al.}(2013)\citenamefont {Xiang},
  \citenamefont {Ashhab}, \citenamefont {You},\ and\ \citenamefont
  {Nori}}]{hybrid_sc}%
  \BibitemOpen
  \bibfield  {author} {\bibinfo {author} {\bibfnamefont {Z.-L.}\ \bibnamefont
  {Xiang}}, \bibinfo {author} {\bibfnamefont {S.}~\bibnamefont {Ashhab}},
  \bibinfo {author} {\bibfnamefont {J.~Q.}\ \bibnamefont {You}}, \ and\
  \bibinfo {author} {\bibfnamefont {F.}~\bibnamefont {Nori}},\ }\href {\doibase
  10.1103/RevModPhys.85.623} {\bibfield  {journal} {\bibinfo  {journal} {Rev.
  Mod. Phys.}\ }\textbf {\bibinfo {volume} {85}},\ \bibinfo {pages} {623}
  (\bibinfo {year} {2013})}\BibitemShut {NoStop}%
\bibitem [{\citenamefont {Gao}\ \emph {et~al.}(2011)\citenamefont {Gao},
  \citenamefont {Liu},\ and\ \citenamefont {Wang}}]{rydberg_om_sc}%
  \BibitemOpen
  \bibfield  {author} {\bibinfo {author} {\bibfnamefont {M.}~\bibnamefont
  {Gao}}, \bibinfo {author} {\bibfnamefont {Y.-x.}\ \bibnamefont {Liu}}, \ and\
  \bibinfo {author} {\bibfnamefont {X.-B.}\ \bibnamefont {Wang}},\ }\href
  {\doibase 10.1103/PhysRevA.83.022309} {\bibfield  {journal} {\bibinfo
  {journal} {Phys. Rev. A}\ }\textbf {\bibinfo {volume} {83}},\ \bibinfo
  {pages} {022309} (\bibinfo {year} {2011})}\BibitemShut {NoStop}%
\bibitem [{\citenamefont {Hafezi}\ \emph {et~al.}(2012)\citenamefont {Hafezi},
  \citenamefont {Kim}, \citenamefont {Rolston}, \citenamefont {Orozco},
  \citenamefont {Lev},\ and\ \citenamefont {Taylor}}]{hafezi_rydberg}%
  \BibitemOpen
  \bibfield  {author} {\bibinfo {author} {\bibfnamefont {M.}~\bibnamefont
  {Hafezi}}, \bibinfo {author} {\bibfnamefont {Z.}~\bibnamefont {Kim}},
  \bibinfo {author} {\bibfnamefont {S.~L.}\ \bibnamefont {Rolston}}, \bibinfo
  {author} {\bibfnamefont {L.~A.}\ \bibnamefont {Orozco}}, \bibinfo {author}
  {\bibfnamefont {B.~L.}\ \bibnamefont {Lev}}, \ and\ \bibinfo {author}
  {\bibfnamefont {J.~M.}\ \bibnamefont {Taylor}},\ }\href {\doibase
  10.1103/PhysRevA.85.020302} {\bibfield  {journal} {\bibinfo  {journal} {Phys.
  Rev. A}\ }\textbf {\bibinfo {volume} {85}},\ \bibinfo {pages} {020302}
  (\bibinfo {year} {2012})}\BibitemShut {NoStop}%
\bibitem [{\citenamefont {Petrosyan}\ and\ \citenamefont
  {Fleischhauer}(2008)}]{petrosyan_rydberg}%
  \BibitemOpen
  \bibfield  {author} {\bibinfo {author} {\bibfnamefont {D.}~\bibnamefont
  {Petrosyan}}\ and\ \bibinfo {author} {\bibfnamefont {M.}~\bibnamefont
  {Fleischhauer}},\ }\href {\doibase 10.1103/PhysRevLett.100.170501} {\bibfield
   {journal} {\bibinfo  {journal} {Phys. Rev. Lett.}\ }\textbf {\bibinfo
  {volume} {100}},\ \bibinfo {pages} {170501} (\bibinfo {year}
  {2008})}\BibitemShut {NoStop}%
\bibitem [{\citenamefont {Petrosyan}\ \emph {et~al.}(2009)\citenamefont
  {Petrosyan}, \citenamefont {Bensky}, \citenamefont {Kurizki}, \citenamefont
  {Mazets}, \citenamefont {Majer},\ and\ \citenamefont
  {Schmiedmayer}}]{petrosyan_rydberg_2}%
  \BibitemOpen
  \bibfield  {author} {\bibinfo {author} {\bibfnamefont {D.}~\bibnamefont
  {Petrosyan}}, \bibinfo {author} {\bibfnamefont {G.}~\bibnamefont {Bensky}},
  \bibinfo {author} {\bibfnamefont {G.}~\bibnamefont {Kurizki}}, \bibinfo
  {author} {\bibfnamefont {I.}~\bibnamefont {Mazets}}, \bibinfo {author}
  {\bibfnamefont {J.}~\bibnamefont {Majer}}, \ and\ \bibinfo {author}
  {\bibfnamefont {J.}~\bibnamefont {Schmiedmayer}},\ }\href {\doibase
  10.1103/PhysRevA.79.040304} {\bibfield  {journal} {\bibinfo  {journal} {Phys.
  Rev. A}\ }\textbf {\bibinfo {volume} {79}},\ \bibinfo {pages} {040304}
  (\bibinfo {year} {2009})}\BibitemShut {NoStop}%
\bibitem [{\citenamefont {Gorshkov}\ \emph {et~al.}(2011)\citenamefont
  {Gorshkov}, \citenamefont {Otterbach}, \citenamefont {Fleischhauer},
  \citenamefont {Pohl},\ and\ \citenamefont {Lukin}}]{Gorshkov2011}%
  \BibitemOpen
  \bibfield  {author} {\bibinfo {author} {\bibfnamefont {A.~V.}\ \bibnamefont
  {Gorshkov}}, \bibinfo {author} {\bibfnamefont {J.}~\bibnamefont {Otterbach}},
  \bibinfo {author} {\bibfnamefont {M.}~\bibnamefont {Fleischhauer}}, \bibinfo
  {author} {\bibfnamefont {T.}~\bibnamefont {Pohl}}, \ and\ \bibinfo {author}
  {\bibfnamefont {M.~D.}\ \bibnamefont {Lukin}},\ }\href {\doibase
  10.1103/PhysRevLett.107.133602} {\bibfield  {journal} {\bibinfo  {journal}
  {Phys. Rev. Lett.}\ }\textbf {\bibinfo {volume} {107}},\ \bibinfo {pages}
  {133602} (\bibinfo {year} {2011})}\BibitemShut {NoStop}%
\bibitem [{\citenamefont {Otterbach}\ \emph {et~al.}(2013)\citenamefont
  {Otterbach}, \citenamefont {Moos}, \citenamefont {Muth},\ and\ \citenamefont
  {Fleischhauer}}]{Otterbach2013}%
  \BibitemOpen
  \bibfield  {author} {\bibinfo {author} {\bibfnamefont {J.}~\bibnamefont
  {Otterbach}}, \bibinfo {author} {\bibfnamefont {M.}~\bibnamefont {Moos}},
  \bibinfo {author} {\bibfnamefont {D.}~\bibnamefont {Muth}}, \ and\ \bibinfo
  {author} {\bibfnamefont {M.}~\bibnamefont {Fleischhauer}},\ }\href {\doibase
  10.1103/PhysRevLett.111.113001} {\bibfield  {journal} {\bibinfo  {journal}
  {Phys. Rev. Lett.}\ }\textbf {\bibinfo {volume} {111}},\ \bibinfo {pages}
  {113001} (\bibinfo {year} {2013})}\BibitemShut {NoStop}%
\bibitem [{\citenamefont {Rabl}\ \emph {et~al.}(2009)\citenamefont {Rabl},
  \citenamefont {Cappellaro}, \citenamefont {Dutt}, \citenamefont {Jiang},
  \citenamefont {Maze},\ and\ \citenamefont {Lukin}}]{rabl09}%
  \BibitemOpen
  \bibfield  {author} {\bibinfo {author} {\bibfnamefont {P.}~\bibnamefont
  {Rabl}}, \bibinfo {author} {\bibfnamefont {P.}~\bibnamefont {Cappellaro}},
  \bibinfo {author} {\bibfnamefont {M.~V.~G.}\ \bibnamefont {Dutt}}, \bibinfo
  {author} {\bibfnamefont {L.}~\bibnamefont {Jiang}}, \bibinfo {author}
  {\bibfnamefont {J.~R.}\ \bibnamefont {Maze}}, \ and\ \bibinfo {author}
  {\bibfnamefont {M.~D.}\ \bibnamefont {Lukin}},\ }\href {\doibase
  10.1103/PhysRevB.79.041302} {\bibfield  {journal} {\bibinfo  {journal} {Phys.
  Rev. B}\ }\textbf {\bibinfo {volume} {79}},\ \bibinfo {pages} {041302}
  (\bibinfo {year} {2009})}\BibitemShut {NoStop}%
\bibitem [{\citenamefont {Bennett}\ \emph {et~al.}(2010)\citenamefont
  {Bennett}, \citenamefont {Cockins}, \citenamefont {Miyahara}, \citenamefont
  {Gr\"utter},\ and\ \citenamefont {Clerk}}]{bennet_em}%
  \BibitemOpen
  \bibfield  {author} {\bibinfo {author} {\bibfnamefont {S.~D.}\ \bibnamefont
  {Bennett}}, \bibinfo {author} {\bibfnamefont {L.}~\bibnamefont {Cockins}},
  \bibinfo {author} {\bibfnamefont {Y.}~\bibnamefont {Miyahara}}, \bibinfo
  {author} {\bibfnamefont {P.}~\bibnamefont {Gr\"utter}}, \ and\ \bibinfo
  {author} {\bibfnamefont {A.~A.}\ \bibnamefont {Clerk}},\ }\href {\doibase
  10.1103/PhysRevLett.104.017203} {\bibfield  {journal} {\bibinfo  {journal}
  {Phys. Rev. Lett.}\ }\textbf {\bibinfo {volume} {104}},\ \bibinfo {pages}
  {017203} (\bibinfo {year} {2010})}\BibitemShut {NoStop}%
\bibitem [{\citenamefont {Rabl}\ \emph {et~al.}(2010)\citenamefont {Rabl},
  \citenamefont {Kolkowitz}, \citenamefont {Koppens}, \citenamefont {Harris},
  \citenamefont {Zoller},\ and\ \citenamefont {Lukin}}]{rabl10}%
  \BibitemOpen
  \bibfield  {author} {\bibinfo {author} {\bibfnamefont {P.}~\bibnamefont
  {Rabl}}, \bibinfo {author} {\bibfnamefont {S.~J.}\ \bibnamefont {Kolkowitz}},
  \bibinfo {author} {\bibfnamefont {F.~H.~L.}\ \bibnamefont {Koppens}},
  \bibinfo {author} {\bibfnamefont {J.~G.~E.}\ \bibnamefont {Harris}}, \bibinfo
  {author} {\bibfnamefont {P.}~\bibnamefont {Zoller}}, \ and\ \bibinfo {author}
  {\bibfnamefont {M.~D.}\ \bibnamefont {Lukin}},\ }\href
  {http://dx.doi.org/10.1038/nphys1679} {\bibfield  {journal} {\bibinfo
  {journal} {Nat Phys}\ }\textbf {\bibinfo {volume} {6}},\ \bibinfo {pages}
  {602} (\bibinfo {year} {2010})}\BibitemShut {NoStop}%
\bibitem [{\citenamefont {Seok}\ \emph {et~al.}(2012)\citenamefont {Seok},
  \citenamefont {Buchmann}, \citenamefont {Singh}, \citenamefont {Steinke},\
  and\ \citenamefont {Meystre}}]{jun_om}%
  \BibitemOpen
  \bibfield  {author} {\bibinfo {author} {\bibfnamefont {H.}~\bibnamefont
  {Seok}}, \bibinfo {author} {\bibfnamefont {L.~F.}\ \bibnamefont {Buchmann}},
  \bibinfo {author} {\bibfnamefont {S.}~\bibnamefont {Singh}}, \bibinfo
  {author} {\bibfnamefont {S.~K.}\ \bibnamefont {Steinke}}, \ and\ \bibinfo
  {author} {\bibfnamefont {P.}~\bibnamefont {Meystre}},\ }\href {\doibase
  10.1103/PhysRevA.85.033822} {\bibfield  {journal} {\bibinfo  {journal} {Phys.
  Rev. A}\ }\textbf {\bibinfo {volume} {85}},\ \bibinfo {pages} {033822}
  (\bibinfo {year} {2012})}\BibitemShut {NoStop}%
\bibitem [{\citenamefont {Singh}\ \emph {et~al.}(2008)\citenamefont {Singh},
  \citenamefont {Bhattacharya}, \citenamefont {Dutta},\ and\ \citenamefont
  {Meystre}}]{swati_om}%
  \BibitemOpen
  \bibfield  {author} {\bibinfo {author} {\bibfnamefont {S.}~\bibnamefont
  {Singh}}, \bibinfo {author} {\bibfnamefont {M.}~\bibnamefont {Bhattacharya}},
  \bibinfo {author} {\bibfnamefont {O.}~\bibnamefont {Dutta}}, \ and\ \bibinfo
  {author} {\bibfnamefont {P.}~\bibnamefont {Meystre}},\ }\href {\doibase
  10.1103/PhysRevLett.101.263603} {\bibfield  {journal} {\bibinfo  {journal}
  {Phys. Rev. Lett.}\ }\textbf {\bibinfo {volume} {101}},\ \bibinfo {pages}
  {263603} (\bibinfo {year} {2008})}\BibitemShut {NoStop}%
\bibitem [{\citenamefont {Kippenberg}\ and\ \citenamefont
  {Vahala}(2008)}]{kip_vah_om}%
  \BibitemOpen
  \bibfield  {author} {\bibinfo {author} {\bibfnamefont {T.~J.}\ \bibnamefont
  {Kippenberg}}\ and\ \bibinfo {author} {\bibfnamefont {K.~J.}\ \bibnamefont
  {Vahala}},\ }\href {\doibase 10.1126/science.1156032} {\bibfield  {journal}
  {\bibinfo  {journal} {Science}\ }\textbf {\bibinfo {volume} {321}},\ \bibinfo
  {pages} {1172} (\bibinfo {year} {2008})}\BibitemShut {NoStop}%
\bibitem [{\citenamefont {Aspelmeyer}\ \emph {et~al.}(2010)\citenamefont
  {Aspelmeyer}, \citenamefont {Gr\"{o}blacher}, \citenamefont {Hammerer},\ and\
  \citenamefont {Kiesel}}]{aspel_om}%
  \BibitemOpen
  \bibfield  {author} {\bibinfo {author} {\bibfnamefont {M.}~\bibnamefont
  {Aspelmeyer}}, \bibinfo {author} {\bibfnamefont {S.}~\bibnamefont
  {Gr\"{o}blacher}}, \bibinfo {author} {\bibfnamefont {K.}~\bibnamefont
  {Hammerer}}, \ and\ \bibinfo {author} {\bibfnamefont {N.}~\bibnamefont
  {Kiesel}},\ }\href {\doibase 10.1364/JOSAB.27.00A189} {\bibfield  {journal}
  {\bibinfo  {journal} {J. Opt. Soc. Am. B}\ }\textbf {\bibinfo {volume}
  {27}},\ \bibinfo {pages} {A189} (\bibinfo {year} {2010})}\BibitemShut
  {NoStop}%
\bibitem [{\citenamefont {Marquardt}\ and\ \citenamefont
  {Girvin}(2009)}]{mar_gir_om}%
  \BibitemOpen
  \bibfield  {author} {\bibinfo {author} {\bibfnamefont {F.}~\bibnamefont
  {Marquardt}}\ and\ \bibinfo {author} {\bibfnamefont {S.~M.}\ \bibnamefont
  {Girvin}},\ }\href {\doibase 10.1103/Physics.2.40} {\bibfield  {journal}
  {\bibinfo  {journal} {Physics}\ }\textbf {\bibinfo {volume} {2}},\ \bibinfo
  {pages} {40} (\bibinfo {year} {2009})}\BibitemShut {NoStop}%
\bibitem [{\citenamefont {Aspelmeyer}\ \emph {et~al.}(2012)\citenamefont
  {Aspelmeyer}, \citenamefont {Meystre},\ and\ \citenamefont
  {Schwab}}]{aspel_mey_schw_om}%
  \BibitemOpen
  \bibfield  {author} {\bibinfo {author} {\bibfnamefont {M.}~\bibnamefont
  {Aspelmeyer}}, \bibinfo {author} {\bibfnamefont {P.}~\bibnamefont {Meystre}},
  \ and\ \bibinfo {author} {\bibfnamefont {K.}~\bibnamefont {Schwab}},\ }\href
  {\doibase 10.1063/PT.3.1640} {\bibfield  {journal} {\bibinfo  {journal}
  {Physics Today}\ }\textbf {\bibinfo {volume} {65}},\ \bibinfo {pages} {29}
  (\bibinfo {year} {2012})}\BibitemShut {NoStop}%
\bibitem [{\citenamefont {Meystre}(2013)}]{meystre_om}%
  \BibitemOpen
  \bibfield  {author} {\bibinfo {author} {\bibfnamefont {P.}~\bibnamefont
  {Meystre}},\ }\href {\doibase 10.1002/andp.201200226} {\bibfield  {journal}
  {\bibinfo  {journal} {Annalen der Physik}\ }\textbf {\bibinfo {volume}
  {525}},\ \bibinfo {pages} {215} (\bibinfo {year} {2013})}\BibitemShut
  {NoStop}%
\bibitem [{\citenamefont {Stamper-Kurn}(2012)}]{stamper_om}%
  \BibitemOpen
  \bibfield  {author} {\bibinfo {author} {\bibfnamefont {D.~M.}\ \bibnamefont
  {Stamper-Kurn}},\ }\href {http://arxiv.org/abs/1204.4351} {\  (\bibinfo
  {year} {2012})},\ \Eprint {http://arxiv.org/abs/arxiv: 1204.4351} {arxiv:
  1204.4351 [quant-ph]} \BibitemShut {NoStop}%
\bibitem [{\citenamefont {Aspelmeyer}\ \emph {et~al.}(2013)\citenamefont
  {Aspelmeyer}, \citenamefont {Kippenberg},\ and\ \citenamefont
  {Marquardt}}]{review_om}%
  \BibitemOpen
  \bibfield  {author} {\bibinfo {author} {\bibfnamefont {M.}~\bibnamefont
  {Aspelmeyer}}, \bibinfo {author} {\bibfnamefont {T.~J.}\ \bibnamefont
  {Kippenberg}}, \ and\ \bibinfo {author} {\bibfnamefont {F.}~\bibnamefont
  {Marquardt}},\ }\href {http://arxiv.org/abs/1303.0733} {\  (\bibinfo {year}
  {2013})},\ \Eprint {http://arxiv.org/abs/arxiv: 1303.0733} {arxiv: 1303.0733
  [cond-mat.mes-hall]} \BibitemShut {NoStop}%
\bibitem [{\citenamefont {Gallagher}(1994)}]{gallagher_book}%
  \BibitemOpen
  \bibfield  {author} {\bibinfo {author} {\bibfnamefont {T.}~\bibnamefont
  {Gallagher}},\ }\href@noop {} {\emph {\bibinfo {title} {Rydberg atoms}}},\
  \bibinfo {edition} {1st}\ ed.\ (\bibinfo  {publisher} {Cambridge University
  Press},\ \bibinfo {address} {Cambridge},\ \bibinfo {year} {1994})\BibitemShut
  {NoStop}%
\bibitem [{Note1()}]{Note1}%
  \BibitemOpen
  \bibinfo {note} {See details in the Supplemental Material which includes
  Refs. \cite {saffmantrap,kuzmichtrap,saffmantrap2, honer_deph, Kubler2013,
  bariani_retrieval, myro_molmer, petrosyan_2013}}\BibitemShut {NoStop}%
\bibitem [{Note2()}]{Note2}%
  \BibitemOpen
  \bibinfo {note} {We use bold symbols to indicate vectors, while the italic
  refers to their moduli}\BibitemShut {NoStop}%
\bibitem [{\citenamefont {Jackson}(1975)}]{jackson}%
  \BibitemOpen
  \bibfield  {author} {\bibinfo {author} {\bibfnamefont {J.}~\bibnamefont
  {Jackson}},\ }\href@noop {} {\emph {\bibinfo {title} {Classical
  Electrodynamics}}},\ \bibinfo {edition} {2nd}\ ed.\ (\bibinfo  {publisher}
  {J. Wiley},\ \bibinfo {address} {New York},\ \bibinfo {year}
  {1975})\BibitemShut {NoStop}%
\bibitem [{\citenamefont {Werlang}\ \emph {et~al.}(2008)\citenamefont
  {Werlang}, \citenamefont {Dodonov}, \citenamefont {Duzzioni},\ and\
  \citenamefont {Villas-B\^oas}}]{beyondRWA}%
  \BibitemOpen
  \bibfield  {author} {\bibinfo {author} {\bibfnamefont {T.}~\bibnamefont
  {Werlang}}, \bibinfo {author} {\bibfnamefont {A.~V.}\ \bibnamefont
  {Dodonov}}, \bibinfo {author} {\bibfnamefont {E.~I.}\ \bibnamefont
  {Duzzioni}}, \ and\ \bibinfo {author} {\bibfnamefont {C.~J.}\ \bibnamefont
  {Villas-B\^oas}},\ }\href {\doibase 10.1103/PhysRevA.78.053805} {\bibfield
  {journal} {\bibinfo  {journal} {Phys. Rev. A}\ }\textbf {\bibinfo {volume}
  {78}},\ \bibinfo {pages} {053805} (\bibinfo {year} {2008})}\BibitemShut
  {NoStop}%
\bibitem [{Note3()}]{Note3}%
  \BibitemOpen
  \bibinfo {note} {See Supplemental Material for a detailed comparison of the
  two terms}\BibitemShut {NoStop}%
\bibitem [{\citenamefont {Saffman}\ \emph {et~al.}(2010)\citenamefont
  {Saffman}, \citenamefont {Walker},\ and\ \citenamefont
  {M{\o}lmer}}]{saffman_review}%
  \BibitemOpen
  \bibfield  {author} {\bibinfo {author} {\bibfnamefont {M.}~\bibnamefont
  {Saffman}}, \bibinfo {author} {\bibfnamefont {T.~G.}\ \bibnamefont {Walker}},
  \ and\ \bibinfo {author} {\bibfnamefont {K.}~\bibnamefont {M{\o}lmer}},\
  }\href {\doibase 10.1103/RevModPhys.82.2313} {\bibfield  {journal} {\bibinfo
  {journal} {Rev. Mod. Phys.}\ }\textbf {\bibinfo {volume} {82}},\ \bibinfo
  {pages} {2313} (\bibinfo {year} {2010})}\BibitemShut {NoStop}%
\bibitem [{\citenamefont {Lukin}\ \emph {et~al.}(2001)\citenamefont {Lukin},
  \citenamefont {Fleischhauer}, \citenamefont {Cote}, \citenamefont {Duan},
  \citenamefont {Jaksch}, \citenamefont {Cirac},\ and\ \citenamefont
  {Zoller}}]{blockade_lukin}%
  \BibitemOpen
  \bibfield  {author} {\bibinfo {author} {\bibfnamefont {M.~D.}\ \bibnamefont
  {Lukin}}, \bibinfo {author} {\bibfnamefont {M.}~\bibnamefont {Fleischhauer}},
  \bibinfo {author} {\bibfnamefont {R.}~\bibnamefont {Cote}}, \bibinfo {author}
  {\bibfnamefont {L.~M.}\ \bibnamefont {Duan}}, \bibinfo {author}
  {\bibfnamefont {D.}~\bibnamefont {Jaksch}}, \bibinfo {author} {\bibfnamefont
  {J.~I.}\ \bibnamefont {Cirac}}, \ and\ \bibinfo {author} {\bibfnamefont
  {P.}~\bibnamefont {Zoller}},\ }\href {\doibase 10.1103/PhysRevLett.87.037901}
  {\bibfield  {journal} {\bibinfo  {journal} {Phys. Rev. Lett.}\ }\textbf
  {\bibinfo {volume} {87}},\ \bibinfo {pages} {037901} (\bibinfo {year}
  {2001})}\BibitemShut {NoStop}%
\bibitem [{\citenamefont {Tong}\ \emph {et~al.}(2004)\citenamefont {Tong},
  \citenamefont {Farooqi}, \citenamefont {Stanojevic}, \citenamefont
  {Krishnan}, \citenamefont {Zhang}, \citenamefont {C\^ot\'e}, \citenamefont
  {Eyler},\ and\ \citenamefont {Gould}}]{blockade_tong}%
  \BibitemOpen
  \bibfield  {author} {\bibinfo {author} {\bibfnamefont {D.}~\bibnamefont
  {Tong}}, \bibinfo {author} {\bibfnamefont {S.~M.}\ \bibnamefont {Farooqi}},
  \bibinfo {author} {\bibfnamefont {J.}~\bibnamefont {Stanojevic}}, \bibinfo
  {author} {\bibfnamefont {S.}~\bibnamefont {Krishnan}}, \bibinfo {author}
  {\bibfnamefont {Y.~P.}\ \bibnamefont {Zhang}}, \bibinfo {author}
  {\bibfnamefont {R.}~\bibnamefont {C\^ot\'e}}, \bibinfo {author}
  {\bibfnamefont {E.~E.}\ \bibnamefont {Eyler}}, \ and\ \bibinfo {author}
  {\bibfnamefont {P.~L.}\ \bibnamefont {Gould}},\ }\href {\doibase
  10.1103/PhysRevLett.93.063001} {\bibfield  {journal} {\bibinfo  {journal}
  {Phys. Rev. Lett.}\ }\textbf {\bibinfo {volume} {93}},\ \bibinfo {pages}
  {063001} (\bibinfo {year} {2004})}\BibitemShut {NoStop}%
\bibitem [{\citenamefont {Dudin}\ \emph {et~al.}(2012)\citenamefont {Dudin},
  \citenamefont {Li}, \citenamefont {Bariani},\ and\ \citenamefont
  {Kuzmich}}]{kuzmich_blockade}%
  \BibitemOpen
  \bibfield  {author} {\bibinfo {author} {\bibfnamefont {Y.~O.}\ \bibnamefont
  {Dudin}}, \bibinfo {author} {\bibfnamefont {L.}~\bibnamefont {Li}}, \bibinfo
  {author} {\bibfnamefont {F.}~\bibnamefont {Bariani}}, \ and\ \bibinfo
  {author} {\bibfnamefont {A.}~\bibnamefont {Kuzmich}},\ }\href
  {http://dx.doi.org/10.1038/nphys2413} {\bibfield  {journal} {\bibinfo
  {journal} {Nat Phys}\ }\textbf {\bibinfo {volume} {8}},\ \bibinfo {pages}
  {790} (\bibinfo {year} {2012})}\BibitemShut {NoStop}%
\bibitem [{Note4()}]{Note4}%
  \BibitemOpen
  \bibinfo {note} {See details and references in the Supplemental
  Material.}\BibitemShut {Stop}%
\bibitem [{\citenamefont {Liao}\ \emph {et~al.}(2010)\citenamefont {Liao},
  \citenamefont {Li}, \citenamefont {Hishita},\ and\ \citenamefont
  {Koide}}]{diamondcantilever}%
  \BibitemOpen
  \bibfield  {author} {\bibinfo {author} {\bibfnamefont {M.}~\bibnamefont
  {Liao}}, \bibinfo {author} {\bibfnamefont {C.}~\bibnamefont {Li}}, \bibinfo
  {author} {\bibfnamefont {S.}~\bibnamefont {Hishita}}, \ and\ \bibinfo
  {author} {\bibfnamefont {Y.}~\bibnamefont {Koide}},\ }\href@noop {}
  {\bibfield  {journal} {\bibinfo  {journal} {Journal of Micromechanics and
  Microengineering}\ }\textbf {\bibinfo {volume} {20}},\ \bibinfo {pages}
  {085002} (\bibinfo {year} {2010})}\BibitemShut {NoStop}%
\bibitem [{\citenamefont {Burek}\ \emph {et~al.}(2012)\citenamefont {Burek},
  \citenamefont {de~Leon}, \citenamefont {Shields}, \citenamefont {Hausmann},
  \citenamefont {Chu}, \citenamefont {Quan}, \citenamefont {Zibrov},
  \citenamefont {Park}, \citenamefont {Lukin},\ and\ \citenamefont {Lon{\v
  c}ar}}]{diamondcantilever2}%
  \BibitemOpen
  \bibfield  {author} {\bibinfo {author} {\bibfnamefont {M.~J.}\ \bibnamefont
  {Burek}}, \bibinfo {author} {\bibfnamefont {N.~P.}\ \bibnamefont {de~Leon}},
  \bibinfo {author} {\bibfnamefont {B.~J.}\ \bibnamefont {Shields}}, \bibinfo
  {author} {\bibfnamefont {B.~J.~M.}\ \bibnamefont {Hausmann}}, \bibinfo
  {author} {\bibfnamefont {Y.}~\bibnamefont {Chu}}, \bibinfo {author}
  {\bibfnamefont {Q.}~\bibnamefont {Quan}}, \bibinfo {author} {\bibfnamefont
  {A.~S.}\ \bibnamefont {Zibrov}}, \bibinfo {author} {\bibfnamefont
  {H.}~\bibnamefont {Park}}, \bibinfo {author} {\bibfnamefont {M.~D.}\
  \bibnamefont {Lukin}}, \ and\ \bibinfo {author} {\bibfnamefont
  {M.}~\bibnamefont {Lon{\v c}ar}},\ }\href@noop {} {\bibfield  {journal}
  {\bibinfo  {journal} {Nano Letters}\ }\textbf {\bibinfo {volume} {12}},\
  \bibinfo {pages} {6084} (\bibinfo {year} {2012})}\BibitemShut {NoStop}%
\bibitem [{\citenamefont {Sidles}\ \emph {et~al.}(1995)\citenamefont {Sidles},
  \citenamefont {Garbini}, \citenamefont {Bruland}, \citenamefont {Rugar},
  \citenamefont {Z\"uger}, \citenamefont {Hoen},\ and\ \citenamefont
  {Yannoni}}]{rmpcantilever}%
  \BibitemOpen
  \bibfield  {author} {\bibinfo {author} {\bibfnamefont {J.~A.}\ \bibnamefont
  {Sidles}}, \bibinfo {author} {\bibfnamefont {J.~L.}\ \bibnamefont {Garbini}},
  \bibinfo {author} {\bibfnamefont {K.~J.}\ \bibnamefont {Bruland}}, \bibinfo
  {author} {\bibfnamefont {D.}~\bibnamefont {Rugar}}, \bibinfo {author}
  {\bibfnamefont {O.}~\bibnamefont {Z\"uger}}, \bibinfo {author} {\bibfnamefont
  {S.}~\bibnamefont {Hoen}}, \ and\ \bibinfo {author} {\bibfnamefont {C.~S.}\
  \bibnamefont {Yannoni}},\ }\href {\doibase 10.1103/RevModPhys.67.249}
  {\bibfield  {journal} {\bibinfo  {journal} {Rev. Mod. Phys.}\ }\textbf
  {\bibinfo {volume} {67}},\ \bibinfo {pages} {249} (\bibinfo {year}
  {1995})}\BibitemShut {NoStop}%
\bibitem [{\citenamefont {Kaulakys}(1995)}]{kaulakys}%
  \BibitemOpen
  \bibfield  {author} {\bibinfo {author} {\bibfnamefont {B.}~\bibnamefont
  {Kaulakys}},\ }\href@noop {} {\bibfield  {journal} {\bibinfo  {journal} {J.
  Phys B: At. Mol. Opt. Phys.}\ }\textbf {\bibinfo {volume} {28}},\ \bibinfo
  {pages} {4963} (\bibinfo {year} {1995})}\BibitemShut {NoStop}%
\bibitem [{\citenamefont {Carmichael}(2002)}]{carmicheal}%
  \BibitemOpen
  \bibfield  {author} {\bibinfo {author} {\bibfnamefont {H.}~\bibnamefont
  {Carmichael}},\ }\href@noop {} {\emph {\bibinfo {title} {Statistical Methods
  in Quantum Optics I}}},\ \bibinfo {edition} {2nd}\ ed.\ (\bibinfo
  {publisher} {Springer},\ \bibinfo {year} {2002})\BibitemShut {NoStop}%
\bibitem [{Note5()}]{Note5}%
  \BibitemOpen
  \bibinfo {note} {Explicit form of the equation of motion may be found in the
  Supplemental Material}\BibitemShut {NoStop}%
\bibitem [{\citenamefont {Law}\ and\ \citenamefont
  {Eberly}(1996)}]{law_eberly}%
  \BibitemOpen
  \bibfield  {author} {\bibinfo {author} {\bibfnamefont {C.~K.}\ \bibnamefont
  {Law}}\ and\ \bibinfo {author} {\bibfnamefont {J.~H.}\ \bibnamefont
  {Eberly}},\ }\href {\doibase 10.1103/PhysRevLett.76.1055} {\bibfield
  {journal} {\bibinfo  {journal} {Phys. Rev. Lett.}\ }\textbf {\bibinfo
  {volume} {76}},\ \bibinfo {pages} {1055} (\bibinfo {year}
  {1996})}\BibitemShut {NoStop}%
\bibitem [{\citenamefont {Joshi}\ \emph {et~al.}(2010)\citenamefont {Joshi},
  \citenamefont {Hutter}, \citenamefont {Zimmer}, \citenamefont {Jonson},
  \citenamefont {Andersson},\ and\ \citenamefont {\"Ohberg}}]{joshi10}%
  \BibitemOpen
  \bibfield  {author} {\bibinfo {author} {\bibfnamefont {C.}~\bibnamefont
  {Joshi}}, \bibinfo {author} {\bibfnamefont {A.}~\bibnamefont {Hutter}},
  \bibinfo {author} {\bibfnamefont {F.~E.}\ \bibnamefont {Zimmer}}, \bibinfo
  {author} {\bibfnamefont {M.}~\bibnamefont {Jonson}}, \bibinfo {author}
  {\bibfnamefont {E.}~\bibnamefont {Andersson}}, \ and\ \bibinfo {author}
  {\bibfnamefont {P.}~\bibnamefont {\"Ohberg}},\ }\href {\doibase
  10.1103/PhysRevA.82.043846} {\bibfield  {journal} {\bibinfo  {journal} {Phys.
  Rev. A}\ }\textbf {\bibinfo {volume} {82}},\ \bibinfo {pages} {043846}
  (\bibinfo {year} {2010})}\BibitemShut {NoStop}%
\bibitem [{\citenamefont {Zhang}\ \emph {et~al.}(2011)\citenamefont {Zhang},
  \citenamefont {Robicheaux},\ and\ \citenamefont {Saffman}}]{saffmantrap}%
  \BibitemOpen
  \bibfield  {author} {\bibinfo {author} {\bibfnamefont {S.}~\bibnamefont
  {Zhang}}, \bibinfo {author} {\bibfnamefont {F.}~\bibnamefont {Robicheaux}}, \
  and\ \bibinfo {author} {\bibfnamefont {M.}~\bibnamefont {Saffman}},\ }\href
  {\doibase 10.1103/PhysRevA.84.043408} {\bibfield  {journal} {\bibinfo
  {journal} {Phys. Rev. A}\ }\textbf {\bibinfo {volume} {84}},\ \bibinfo
  {pages} {043408} (\bibinfo {year} {2011})}\BibitemShut {NoStop}%
\bibitem [{\citenamefont {Li}\ \emph {et~al.}(2013)\citenamefont {Li},
  \citenamefont {Dudin},\ and\ \citenamefont {Kuzmich}}]{kuzmichtrap}%
  \BibitemOpen
  \bibfield  {author} {\bibinfo {author} {\bibfnamefont {L.}~\bibnamefont
  {Li}}, \bibinfo {author} {\bibfnamefont {Y.~O.}\ \bibnamefont {Dudin}}, \
  and\ \bibinfo {author} {\bibfnamefont {A.}~\bibnamefont {Kuzmich}},\ }\href
  {http://dx.doi.org/10.1038/nature12227} {\bibfield  {journal} {\bibinfo
  {journal} {Nature}\ }\textbf {\bibinfo {volume} {498}},\ \bibinfo {pages}
  {466} (\bibinfo {year} {2013})}\BibitemShut {NoStop}%
\bibitem [{\citenamefont {Piotrowicz}\ \emph {et~al.}(2013)\citenamefont
  {Piotrowicz}, \citenamefont {Lichtman}, \citenamefont {Maller}, \citenamefont
  {Li}, \citenamefont {Zhang}, \citenamefont {Isenhower},\ and\ \citenamefont
  {Saffman}}]{saffmantrap2}%
  \BibitemOpen
  \bibfield  {author} {\bibinfo {author} {\bibfnamefont {M.~J.}\ \bibnamefont
  {Piotrowicz}}, \bibinfo {author} {\bibfnamefont {M.}~\bibnamefont
  {Lichtman}}, \bibinfo {author} {\bibfnamefont {K.}~\bibnamefont {Maller}},
  \bibinfo {author} {\bibfnamefont {G.}~\bibnamefont {Li}}, \bibinfo {author}
  {\bibfnamefont {S.}~\bibnamefont {Zhang}}, \bibinfo {author} {\bibfnamefont
  {L.}~\bibnamefont {Isenhower}}, \ and\ \bibinfo {author} {\bibfnamefont
  {M.}~\bibnamefont {Saffman}},\ }\href {http://arxiv.org/abs/1305.6102} {\
  (\bibinfo {year} {2013})},\ \Eprint {http://arxiv.org/abs/arxiv: 1305.6102}
  {arxiv: 1305.6102 [physics.atom-ph]} \BibitemShut {NoStop}%
\bibitem [{\citenamefont {Honer}\ \emph {et~al.}(2011)\citenamefont {Honer},
  \citenamefont {L\"ow}, \citenamefont {Weimer}, \citenamefont {Pfau},\ and\
  \citenamefont {B\"uchler}}]{honer_deph}%
  \BibitemOpen
  \bibfield  {author} {\bibinfo {author} {\bibfnamefont {J.}~\bibnamefont
  {Honer}}, \bibinfo {author} {\bibfnamefont {R.}~\bibnamefont {L\"ow}},
  \bibinfo {author} {\bibfnamefont {H.}~\bibnamefont {Weimer}}, \bibinfo
  {author} {\bibfnamefont {T.}~\bibnamefont {Pfau}}, \ and\ \bibinfo {author}
  {\bibfnamefont {H.~P.}\ \bibnamefont {B\"uchler}},\ }\href {\doibase
  10.1103/PhysRevLett.107.093601} {\bibfield  {journal} {\bibinfo  {journal}
  {Phys. Rev. Lett.}\ }\textbf {\bibinfo {volume} {107}},\ \bibinfo {pages}
  {093601} (\bibinfo {year} {2011})}\BibitemShut {NoStop}%
\bibitem [{\citenamefont {K{\"u}bler}\ \emph {et~al.}(2013)\citenamefont
  {K{\"u}bler}, \citenamefont {Booth}, \citenamefont {Sedlacek}, \citenamefont
  {Zabawa},\ and\ \citenamefont {Shaffer}}]{Kubler2013}%
  \BibitemOpen
  \bibfield  {author} {\bibinfo {author} {\bibfnamefont {H.}~\bibnamefont
  {K{\"u}bler}}, \bibinfo {author} {\bibfnamefont {D.}~\bibnamefont {Booth}},
  \bibinfo {author} {\bibfnamefont {J.}~\bibnamefont {Sedlacek}}, \bibinfo
  {author} {\bibfnamefont {P.}~\bibnamefont {Zabawa}}, \ and\ \bibinfo {author}
  {\bibfnamefont {J.}~\bibnamefont {Shaffer}},\ }\href
  {http://arxiv.org/abs/1304.7266} {\  (\bibinfo {year} {2013})},\ \Eprint
  {http://arxiv.org/abs/arxiv: 1304.7266} {arxiv: 1304.7266 [physics.atom-ph]}
  \BibitemShut {NoStop}%
\bibitem [{\citenamefont {Bariani}\ and\ \citenamefont
  {Kennedy}(2012)}]{bariani_retrieval}%
  \BibitemOpen
  \bibfield  {author} {\bibinfo {author} {\bibfnamefont {F.}~\bibnamefont
  {Bariani}}\ and\ \bibinfo {author} {\bibfnamefont {T.~A.~B.}\ \bibnamefont
  {Kennedy}},\ }\href {\doibase 10.1103/PhysRevA.85.033811} {\bibfield
  {journal} {\bibinfo  {journal} {Phys. Rev. A}\ }\textbf {\bibinfo {volume}
  {85}},\ \bibinfo {pages} {033811} (\bibinfo {year} {2012})}\BibitemShut
  {NoStop}%
\bibitem [{\citenamefont {Miroshnychenko}\ \emph {et~al.}(2013)\citenamefont
  {Miroshnychenko}, \citenamefont {Poulsen},\ and\ \citenamefont
  {M\o{}lmer}}]{myro_molmer}%
  \BibitemOpen
  \bibfield  {author} {\bibinfo {author} {\bibfnamefont {Y.}~\bibnamefont
  {Miroshnychenko}}, \bibinfo {author} {\bibfnamefont {U.~V.}\ \bibnamefont
  {Poulsen}}, \ and\ \bibinfo {author} {\bibfnamefont {K.}~\bibnamefont
  {M\o{}lmer}},\ }\href {\doibase 10.1103/PhysRevA.87.023821} {\bibfield
  {journal} {\bibinfo  {journal} {Phys. Rev. A}\ }\textbf {\bibinfo {volume}
  {87}},\ \bibinfo {pages} {023821} (\bibinfo {year} {2013})}\BibitemShut
  {NoStop}%
\bibitem [{\citenamefont {Petrosyan}\ \emph {et~al.}(2013)\citenamefont
  {Petrosyan}, \citenamefont {H\"oning},\ and\ \citenamefont
  {Fleischhauer}}]{petrosyan_2013}%
  \BibitemOpen
  \bibfield  {author} {\bibinfo {author} {\bibfnamefont {D.}~\bibnamefont
  {Petrosyan}}, \bibinfo {author} {\bibfnamefont {M.}~\bibnamefont {H\"oning}},
  \ and\ \bibinfo {author} {\bibfnamefont {M.}~\bibnamefont {Fleischhauer}},\
  }\href {\doibase 10.1103/PhysRevA.87.053414} {\bibfield  {journal} {\bibinfo
  {journal} {Phys. Rev. A}\ }\textbf {\bibinfo {volume} {87}},\ \bibinfo
  {pages} {053414} (\bibinfo {year} {2013})}\BibitemShut {NoStop}%
\end{thebibliography}%
 \newpage
\appendix

\section{Supplemental Material}

\subsection{Rydberg Trapping and Blockade}
The execution of the proposed quantum protocols require the atom to be in the Rydberg level for some time. Rydberg states exhibit a negative polarizability as opposed to atomic ground states and hence require a different trap. For applications that involve single atoms or that are fast compared to the motional broadening of an ensemble, the trap is simply switched off during the Rydberg excitation and the atoms are trapped again once they go back to the ground state.  Recently, different techniques have been proposed and implemented to trap the Rydberg excited levels as well as the ground state \cite{saffmantrap,*kuzmichtrap,*saffmantrap2}. The trapping lifetime already achieved in experiments allows the realization of the protocols discussed in the main text.

In the main text we discuss the cases of single atom and atomic cloud (super-atom) together. In fact, Rydberg blockade of multiple excitation in the case of an ensemble permits to eliminate all the collective states that are far detuned due to strong many-body interactions \cite{saffman_review}. We stress that although equivalent in this problem, single atom and super-atom present a different microscopic structure, which becomes relevant in other applications. For example, the collective nature of the blockade effect is crucial in dissipative effects, such as collective dephasing~\cite{honer_deph,Kubler2013}, super radiant decay~\cite{bariani_retrieval,myro_molmer}, and steady state of strongly driven ensembles~\cite{petrosyan_2013}.

\subsection{Relative strength of dipole couplings}
The dynamical effect arising from $V_{12}$ in the main text is the exchange of phonons between the two cantilevers with rate 
\begin{equation}
\hbar \mathcal{G}_{12} = \frac{Q^2 x_{zp,1} x_{zp,2}}{32 \pi \epsilon_0 R^3}.
\end{equation}
If $\omega_1 = \omega_2$, this resonant effect competes with the atom-cantilever couplings and the relative strength is 
\begin{equation}
\frac{\mathcal{G}_{12}}{\mathcal{G}_{1}} = \frac{1}{8} \, \frac{Q x_{zp,1}}{\mu_{sp}}.
\end{equation} 
We can estimate $\mu_{sp} \sim n^2 e a_0$ where $n$ is the principal quantum number of the Rydberg levels, $e$ is the elementary charge and $a_0$ the Bohr radius. In general $x_{zp} \ll a_0$ and $Q \ll n^2 e$, thus this term is suppressed on the interaction time we are interested in. For the non-resonant case $\omega_1 \neq \omega_2$, it is further suppressed proportionally to the energy detuning $\omega_{12} = \omega_1 - \omega_2$.

\subsection{Master equation for cantilever cooling}
In order to simulate the cooling of the cantilever via the coupling with the Rydberg excitation we use the master equation
\begin{equation}
\frac{\partial \rho}{\partial t} = \frac{i}{\hbar} [H,\rho] + \mathcal{L}(\rho),
\end{equation}
where $\rho$ is the density matrix of the system, $H = H_c + H_{at} + V$ is the Hamiltonian for the atom-cantilever system (see main text) and the Linblad operator for the density matrix is given by:
\begin{align}
\mathcal{L}/\hbar = & \frac{\Gamma_e}{2} (2 \sigma_{ge} \rho \sigma_{eg} -  \sigma_{ee} \rho -  \rho \sigma_{ee}) \nonumber \\ 
                               & + \frac{\gamma_m}{2} (n_{th} + 1) (2 b \rho b^{\dagger} - b^{\dagger}b \rho -  \rho b^{\dagger}b) \nonumber \\ 
                               & + \frac{\gamma_m}{2} n_{th} (2 b^{\dagger} \rho b - bb^{\dagger} \rho -  \rho bb^{\dagger}).
\end{align}
Here 
\begin{equation}
n_{th} = \frac{1}{e^{\frac{\hbar \omega_m}{K_{B}T}}-1},
\end{equation} 
is the average thermal occupation of the cantilever at temperature $T$. The final effective temperature $T_\text{eff}$ is determined from the steady state population of the phononic ground state,
\begin{equation}
P_0 = {1 - e^{-\frac{\hbar \omega_m}{K_{B}T_\text{eff}}}}.
\end{equation}
A simple way to estimate the cooling rate is based on the resonant coherent coupling of the three levels $\ket{g}$, $\ket{p}$ and $\ket{s}$. 
For $\Omega_L \gg \mathcal{G}$, the population are: $\langle\sigma_{gg}\rangle = \langle\sigma_{pp}\rangle \sim 1/2$ and $\langle\sigma_{ss}\rangle \sim \mathcal{G}^2/\Omega_L^2$. The populations oscillate with frequency $\Omega = \sqrt{\mathcal{G}^2 + \Omega_L^2} \sim \Omega_L$.
If $\Omega_R \gg \mathcal{G}$, we may assume that every time we populate the level $\ket{s}$ the atom is quickly transferred into the ground state and a phonon is subtracted from the cantilever. It is thus straightforward to obtain the cooling rate: $\Gamma_{at} \sim \mathcal{G}^2/\Omega_L$.

\end{document}